\documentclass[reprint,amsmath, amssymb, pra, longbibliography, nofootinbib]{revtex4-1}
\usepackage[colorlinks]{hyperref}

\usepackage[colorlinks]{hyperref}   
\usepackage{qcircuit}

\def\exp{\hbox{\rm exp}}
\smallskip 
\def\<{{\langle }}
\def\>{{\rangle }}

\def\ket#1{|#1\rangle}
\def\bra#1{\langle#1|}
\def\kket#1{|#1\rangle\rangle}
\def\bbra#1{\langle\langle#1 |}

\def\exp{\hbox{\rm exp}}
\smallskip
\def\<{{\langle }}
\def\>{{\rangle }}

\usepackage{color}
\usepackage{graphicx}
\usepackage{dcolumn}
\usepackage{bm}

\begin{document}

\title{Parameterized process characterization with reduced resource requirements}

\thanks{This manuscript has been authored by UT-Battelle, LLC, under Contract No. DE-AC0500OR22725 with the U.S. Department of Energy. The United States Government retains and the publisher, by accepting the article for publication, acknowledges that the United States Government retains a non-exclusive, paid-up, irrevocable, world-wide license to publish or reproduce the published form of this manuscript, or allow others to do so, for the United States Government purposes. The Department of Energy will provide public access to these results of federally sponsored research in accordance with the DOE Public Access Plan.}%

\author{ Vicente Leyton-Ortega} 
\email[Corresponding author: ]{leytonorteva@ornl.gov}
 \affiliation{Computational Sciences and Engineering Division, Oak Ridge National Laboratory, Oak Ridge, TN 37831, USA}
 
\author{ Tyler Kharazi} 
 \affiliation{Computational Sciences and Engineering Division, Oak Ridge National Laboratory, Oak Ridge, TN 37831, USA}
 
 \author{ Raphael C. Pooser} 
 \affiliation{Computational Sciences and Engineering Division, Oak Ridge National Laboratory, Oak Ridge, TN 37831, USA}

\date{\today}

\begin{abstract}
Quantum Process Tomography (QPT) is a powerful tool to characterize quantum operations, but it requires considerable resources making it impractical for more than 2-qubit systems. This work proposes an alternative approach that requires significantly fewer resources for {unitary} processes characterization {without prior knowledge of the process} and provides a built-in method for state preparation and measurement (SPAM) error mitigation. By measuring the quantum process as rotated through the $X$ and $Y$ axes on the Bloch Sphere, we can acquire enough information to reconstruct the quantum process matrix $\chi$ and measure its fidelity. We test the algorithm's performance against standard QPT using simulated and physical experiments on several IBM quantum processors and compare the resulting process matrices. We demonstrate in numerical experiments that the method can improve gate fidelity via a noise reduction in the imaginary part of the process matrix, along with a stark decrease in the number of experiments needed to perform the characterization. 

\end{abstract}

                              
\maketitle    


\section{Introduction}
Modern quantum computers are marred by noise that limits the computational reach of these devices. 
The sources of this noise are myriad, including initial state preparation errors, noise introduced during the computation via decoherence and gate noise, and imprecise state readout at measurement \cite{Magesan2012}. There has been extensive work to improve quantum processor units (QPUs) at the hardware level with methods that are generally not accessible at the end-user level. Given the current proliferation of noisy intermediate-scale quantum resources, we seek to develop strategies that end-users can use to calibrate a set of qubits on a physical QPU in a cost-effective (less resource-intensive) manner. 
Furthermore, new algorithms to isolate and characterize noise are critical  for quantifying where QPUs need improvement and benchmarking algorithm performance \cite{Eisert2020}.

One such characterization method, quantum tomography, provides a set of tools to characterize the behavior of quantum dynamical processes through a series of measurements on a complete basis, typically the Pauli basis. Standard quantum process tomography (QPT) reconstructs the underlying quantum process $\mathcal{E}$ by performing state tomography on a set of identical quantum states after applying certain quantum operations, i.e., a quantum circuit \cite{NielsenBook, Poyatos1997, Chuang1997}. Through this tomographic reconstruction in state space, one can infer the region where each generated state lies by applying maximum likelihood estimation~\cite{Baumgratz2013}, Bayesian credibility~\cite{BlumeKohout2010, Ferrie2014}, or confidence regions~\cite{BlumeKohout2012,Christandl2012,Wang2019}.

Process tomography is also a key component in noise characterization and noise mitigation for quantum algorithms~\cite{Zhang2020}. Through a large number of circuit evaluations, or shots, one can deploy statistical and numerical methods to recover the underlying { process matrix representation} of $\mathcal{E}$ \cite{Alexander2020, Korotkov2013, Mohseni2008}. However, the resource requirements inhibit the scalability of QPT-based methods on NISQ hardware. For a complete determination of an n-qubit quantum process, one needs to prepare $4^n \times 3^n$ independent circuit executions to specify a quantum process completely (see Appendix \ref{app:qpt} for more details). This resource overhead makes QPT impractical for characterizing processes involving more than a few qubits. For example, the complete characterization of a 3-qubit quantum process requires $12^3 = 1728$ independent experiments, with each experiment repeated many times to gather sufficient statistics. On the publicly available IBM QPUs, i.e., \texttt{IBMQ Bogota}, a user can send at most 900 independent experiments, falling far short of the 1728 required to characterize a 3-qubit process fully. One may send the complete set of experiments in two separate batches of circuits, but this leaves open the possibility that the device may have changed significantly between experimental runs. Without dedicated access to a QPU, process tomography is practically challenging for $n>3$ qubit systems.  {To dodge this problem, ancilla-\cite{Altepeter2003, Leung2003, Ariano2003, Ariano2001} and error-correction-based\cite{Omkar2015a, Omkar2015b, Mohseni2006, Mohseni2007} QPT schemes were introduced. These methods require sophisticated state and measurement preparations. The number of experiments can be reduced even more if some prior information about the process is known using compressed sensing techniques\cite{Flammia2010, Flammia2012}.  However, the success of compressed sensing depends on the accuracy of the rank knowledge given for the quantum process\cite{Rodionov2014, Shabani2011}.  An adaptive measurement technique introduces a way to characterize any unitary process that does not require any prior assumption about the process\cite{Kim2020}, but it requires an optimization routine to adapt the initial states and measurement operators. On the other hand, the resource requirements can be reduced if one does not require a complete characterization of the quantum process; for instance, randomized benchmarking is commonly use to compute gate fidelities on superconducting QPUs \cite{Emerson2005,Knill2008, Magesan2011}. In summary, these methods assume a specific structure: low-rank restrictions \cite{Shabani2011}, two-qubit processes \cite{Govia_2020}, and a unitary structure that only requires measurements of the diagonal elements of the rotated process matrix.   }

{We consider the results from standard QPT as a reference to measure the performance of our method.}
  In addition to the experimental overhead, QPT assumes perfect readout measurement, yet it is highly sensitive to state preparation and measurement (SPAM) errors \cite{Eisert2020, Korotkov2013}. This assumption can lead to underestimating process fidelity by rolling SPAM errors into the same process as the gates one is trying to characterize. Here, we reduce the complexity of QPT without sparsity assumptions while still offering an exponential improvement in resource cost over standard QPT by assuming a very simple noise model consisting of SPAM error and rotation error (see Fig. \ref{fig:budget}). Characterization uses a series of rotations and measurements tailored to limited access quantum chips such as cloud-based IBM quantum devices, which we dub parametrized process characterization (PPC). We further provide a method to unravel SPAM errors from process characterization by fitting the projective measurement of key quantum states generated by the quantum process to a statistical model influenced by SPAM-type errors. The resulting fit parameters then allow us to reconstruct the underlying quantum process.




\section{Methods}
To illustrate the general idea { of PPC}, we first give a one-qubit example that can be extended to a more general case. We wish to characterize some quantum process ${\cal U}: \rho \rightarrow U \rho\, U^\dagger$, with $\rho$ and $U$ as a one-qubit quantum state and unitary operator,  respectively. Without loss of generality, we shall consider the rotation $Y_\theta = \exp [-i \theta \sigma_y/2]$ and assume that rotation is a noiseless unitary operation in the experimental setup. The projective measurement, along the $z$-axis in the Bloch sphere, of the state $Y_\theta U \ket{s}$, for $\ket{s}\in \lbrace  \ket{0}, \ket{1} \rbrace$, reads
\begin{equation}\label{eq:prob0}
	P^{\cal U} _s (\theta) = \left( 
		\begin{array}{c}
		s_{\theta/2}^2 + |{U}_{s,0}|^2 c_\theta - ({ U}_{s,0} {U}_{s,1}^* +  { U}_{s,1} { U}_{s,0}^*) s_\theta \\[2mm]
		s_{\theta/2}^2 + |{U}_{s,1}|^2 c_\theta + ({U}_{s,0} {U}_{s,1}^* +  {U}_{s,1} {U}_{s,0}^*) s_\theta
	   \end{array}
	\right) \ ,
\end{equation}
with ${U}_{kl} = \bra{l} {U} \ket{k}$, $c_\theta = \cos \theta$, and $s_\theta = \sin \theta$.

To account for readout error, we introduce a classical assignment error modeled by the transition matrix: 
\begin{equation}
	T = \left(
		\begin{array}{cc}
		   t_{00} & 1- t_{11} \\[2mm]
		   1-t_{00} & t_{11}
	   \end{array}
	\right),
\end{equation}
with $t_{00}$ and $t_{11}$ as the probabilities of measuring correctly the states $\ket{0}$ and $\ket{1}$, respectively. $T$ is obtained experimentally via calibration measurements. This matrix represents a binary asymmetric channel, i.e., $t_{00} \neq t_{11}$, that maps the original probability distribution to the experimental observation $Q^{\cal U} _s (\theta) = T \cdot P^{\cal U} _s (\theta)$, which can be rewritten as:
\begin{equation} \label{eq:qdist}
	Q^{\cal U}_{s} (\theta) = \left(
		\begin{array}{c}
			1 - t_{11} \\
			1 - t_{00}
		\end{array}
	\right) + |T| P^{\cal U}_s (\theta - \theta_0).
\end{equation} 
This establishes a way to determine the quantum process by fitting $U_{s,0}$ and $U_{s,1}$ with the experimental data $T$ and $Q^{\cal U}_s(\theta)$.  Notably, with the assumption a transition matrix can be describe readout error, and that errors occur only along the direction of rotation, a single parameter in Eq.~\ref{eq:qdist} can be used to fit the data. 
Here $\theta_0$ is an initial phase representing either a state preparation error or a compilation error in the rotation operator $Y_\theta$. In summary, we evaluate the action of the unitary operator $U$ on a set of rotated states $Y_\theta \ket{0}$, and by fitting the model for the measurement output $T\cdot P_s^{\cal U}(\theta)$ to the experimental output $Q_s^{\cal U}(\theta)$ in (see Eq. \ref{eq:qdist}), we can estimate the components for $U$ in the computational basis.

This procedure can be extended to an $n$-qubit system, considering $Y^s_\theta$ as the main rotation; the superscript $s$ stands for the physical qubit where the rotation is applied. As before, it is enough to consider a subset of the $n$-qubit computational basis where the $s$-th qubit is in the ground state, i.e., the set $\lbrace \ket{\varphi_{k,s}} = \ket{k_0, \cdots , k_{2^n}} \in \lbrace \ket{0}, \ket{1} \rbrace^n : k_s=0 \rbrace$, and determine the action of the unitary operator $U$ on different rotations applied on this set, 
\begin{equation}
    \ket{\psi_{k,s}(\theta)} = U \cdot Y^s_\theta \ket{\varphi_{k,s}}
\end{equation}
for different values of $\theta$ in $[-\pi, \pi]$. The projective measurement from these states, $T\cdot P_{k,s}^{\cal U}(\theta)$, can be seen as a function of $\theta$ with parameters given by the components of $U$, i.e., a function $F_{k,s}(\theta; U_{00}, \cdots , U_{(2^n, 2^n)})$. Thus, the fit parameters of this function to the experimental measurement data give the estimate for $U$. Before applying the algorithm, we use a calibration procedure to determine the transition matrix $T$ and the phase correction $\theta_0$ that uses a similar angular sweep (see Section \ref{sec:calibration}).     
\begin{figure}
    \centering
    \includegraphics[width=0.7\linewidth]{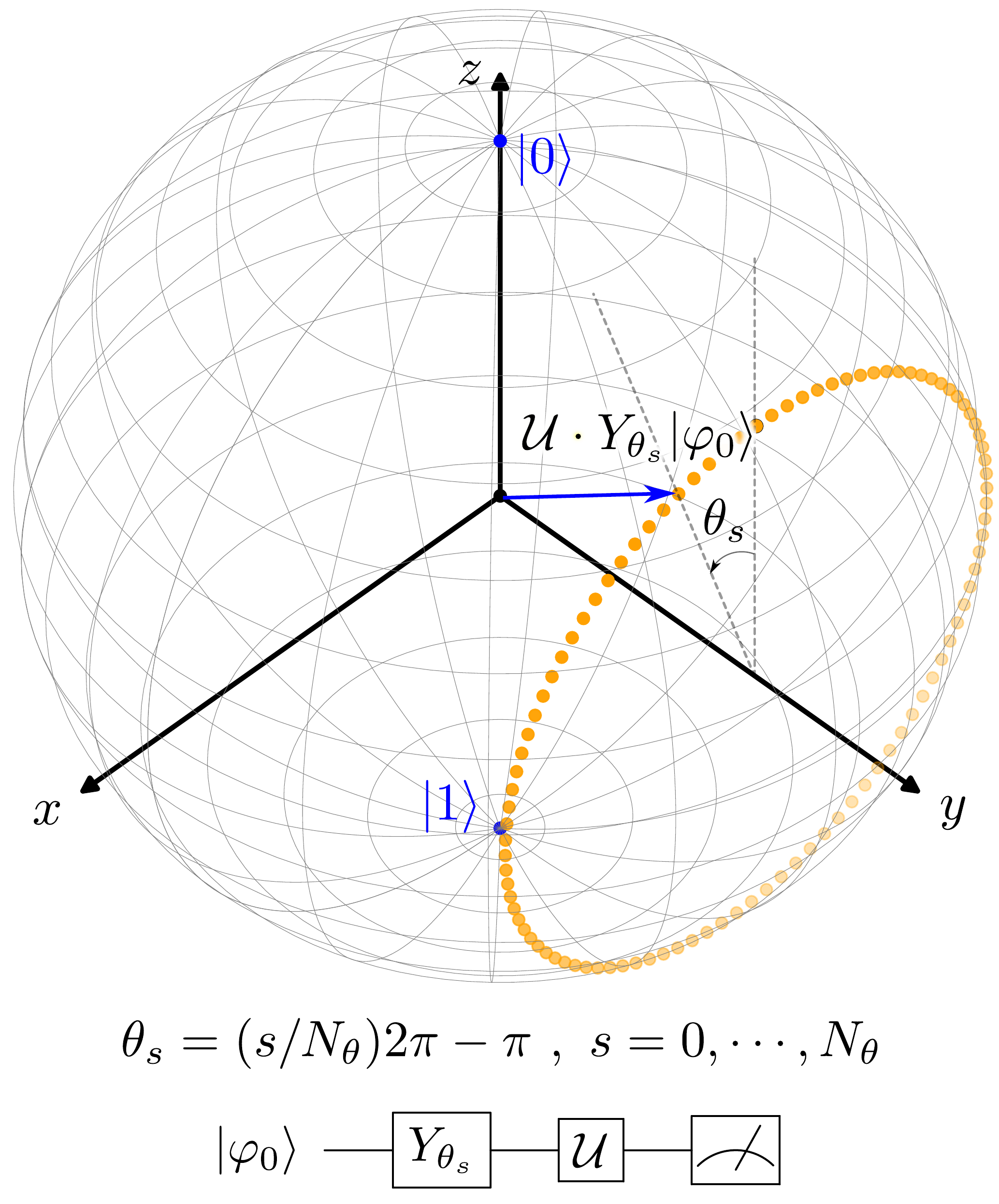}
    \caption{Diagrammatic sketch of PPC in the Bloch sphere for one-qubit systems. With the information obtained from the rotation on the states ${\cal U} \ket{0}$  and ${\cal U} \ket{1}$ we get information about the quantum process represented by $\cal U$. The dotted lines represent the rotations of the states ${\cal U} \ket{0}$ and ${\cal U} \ket{1}$ around the $y$-axis.}
    \label{fig:sketch}
\end{figure}
%


\section{Single Qubit Quantum Process Characterization}\label{sec:QPC}

\begin{figure}  
    \centering
    \includegraphics[width=0.8\linewidth]{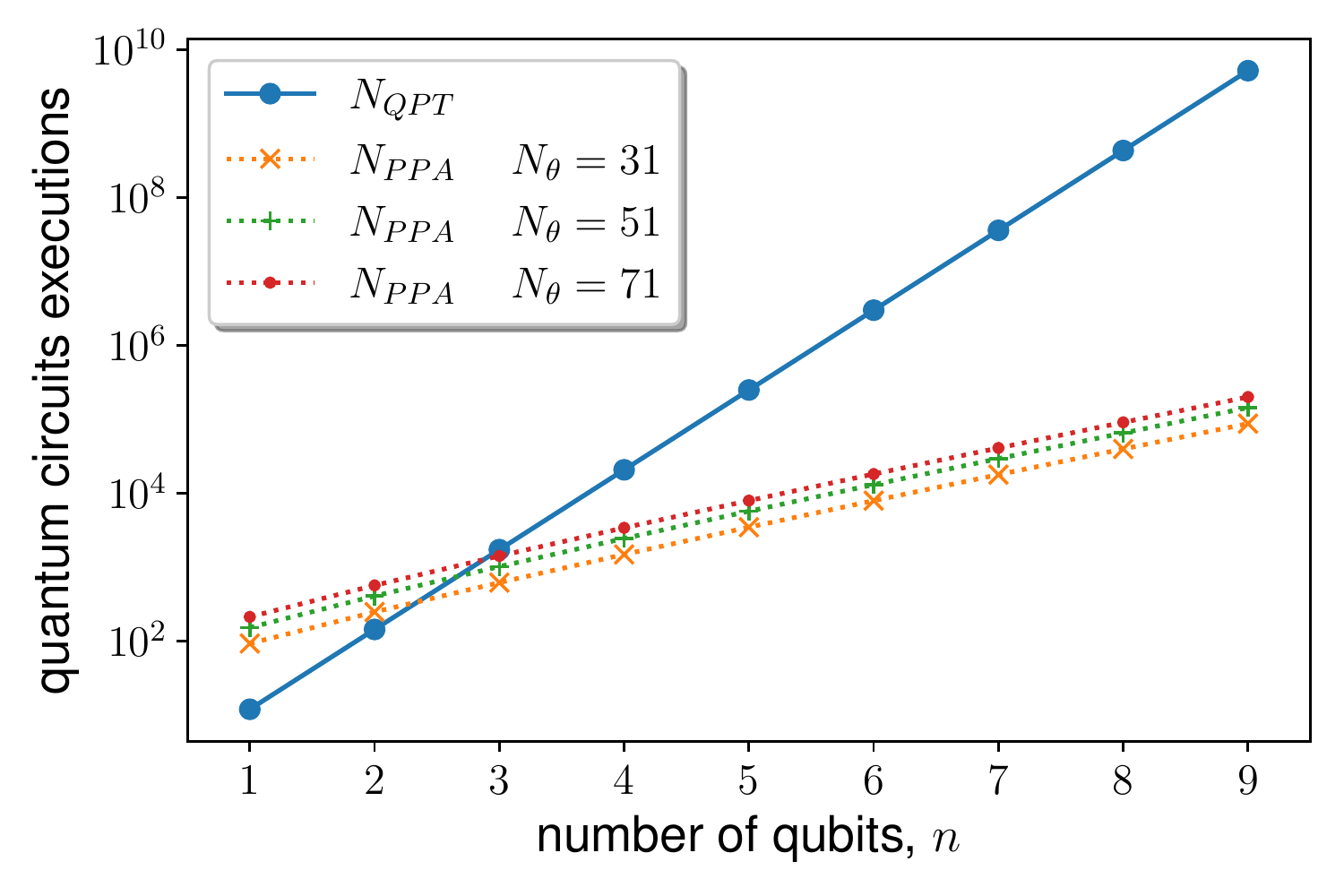}
    \caption{\emph{Number of quantum circuit executions for quantum process characterization:} in this figure, we compare the resources used by QPT against the resources used by PPC. For different numbers of rotations ($N_\theta = 31, 51, $and $71$ lines), QPT surpasses PPC in the number of resources required for its execution for $n >  2 $. }
    \label{fig:budget} 
\end{figure}

For an $n$-qubit quantum process characterization ${\cal U}$, it is necessary to execute $N_{PPC} = 2^{n-1}(n+2) N_\theta$ quantum circuits with $N_\theta$ as the number of rotations the interval $[-\pi,\pi]$ is divided into; $2^{n} N_\theta$ quantum circuits for calibration; and $2^{n-1}n N_\theta$ quantum circuits to obtain data for fitting an estimate of $U$. Both the calibration and the estimation quality depends on $N_\theta$ -a large enough value ensures slight deviations  $(\sigma \sim N_\theta^{-1/2})$ of the model from the experimental data. Note that $N_\theta$ does not depend on the number of qubits. Since QPT scales differently with the number of qubits, $N_{QPT} = 12^n$, PPC with a moderate number of angles becomes a favorable method in cases where $n > 2$, { since $\log (N_{QPT} / N_{{PPC}} ) \sim   (n -1) \log 6  - \log (N_{\theta}/2)$}.  Figure~\ref{fig:budget} shows resource scaling for each protocol.    

We performed simulations and on-hardware experiments for one and two-qubit systems 
using native gates as the target operations to characterize. For these experiments, we set $N_\theta = 51$ rotations and $N = 5000$ experimental repetitions of PPC circuits and employ the Python Symfit \cite{symfit} library to fit the model to the experimental data. Symfit is a Python module that uses symbolic methods to determine the Jacobian in the fitting process analytically. We test our procedure characterizing the X- and H-gate for one-qubit systems and the CX-gate for two-qubit systems applied on different plaquettes on several IBM QPUs. We use $Y_\theta$ for the calibration and characterization since this rotation has lower fidelity than $X_\theta$ (see appendix \ref{app:exp}), and thus constitutes a good benchmark for the method. For the QPT experiments, we use the tomography module of Qiskit Ignis~\cite{Qiskit}.

{ A common way to describe the quantum process $\cal U$ is through the process matrix ${\chi}_{\cal}$ defined as 
\begin{equation} \label{eq:defchi}
 {\cal U} ( \rho) = \sum_{kl} \chi_{kl} P_{k} \rho P_{l} , 
\end{equation}
with $\rho$ as a one qubit state, and $P_{k} \in \lbrace \mathbb{I} , \sigma_{x}, \sigma_{y}, \sigma_{z} \rbrace$ (Pauli basis). It is simple to find the process matrix ${\chi}$, we need to find the representation of $U$ in the Pauli basis and compare $U \rho_{0} U^{\dagger}$ with Eq. \eqref{eq:defchi}, which gives $\chi_{kl} = u_{k}u^{*}_{l} $ with $ u_{k} = {\rm tr} \lbrace {\cal U} P_{k}\rbrace /2$.
}

\begin{figure}[h]
    \centering
    \includegraphics[width=\linewidth]{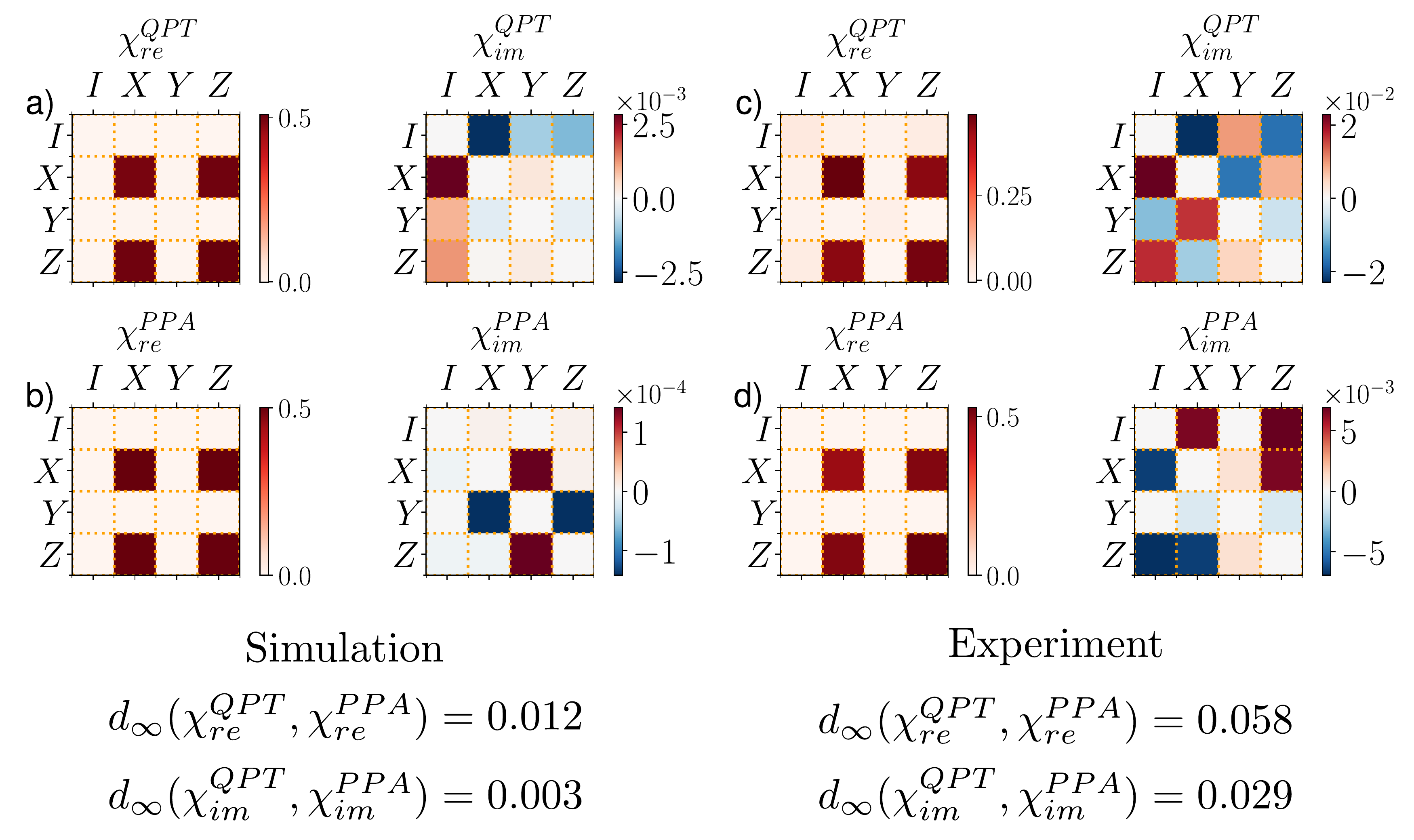}
    \caption{ \emph{Comparison between tomography results for the one-qubit system}. Here we present the $\chi$-matrix representation in the Pauli matrices for the Hadamard gate $\rm H$ using QPT and PPC. In a) and b), we depict the $\chi$-matrix using QPT and PPC, respectively, by running the experiment in a noiseless simulator. In c) and d), we show the resulting $\chi$ matrices, by QPT and PPC, respectively, by running the experiment on the 0th qubit on the IBMQ-Bogota backend. In both cases, we considered $N=5000$ shots and least square estimation for QPT, and $N = 5000$ shots and $N_\theta = 51$ for PPC. There is a good agreement in the $\chi$-matrix's real part, but a difference in the imaginary part, even when using the noiseless simulator. {In the noiseless simulator, the imaginary part is statistically irrelevant since the statistical error is in the order of $N^{-1/2}/2 \sim 7\times 10^{-3} $\cite{Knips2015}. On the other hand, in the QPT experiments, the imaginary part is statistically relevant due to SPAM errors. At the bottom, we present the $d_{\infty}(\cdot)$ distance between the resulting process matrices.}}
    \label{fig:h_char}
\end{figure}

\begin{figure}[ht]
    \centering
    \includegraphics[width=\linewidth]{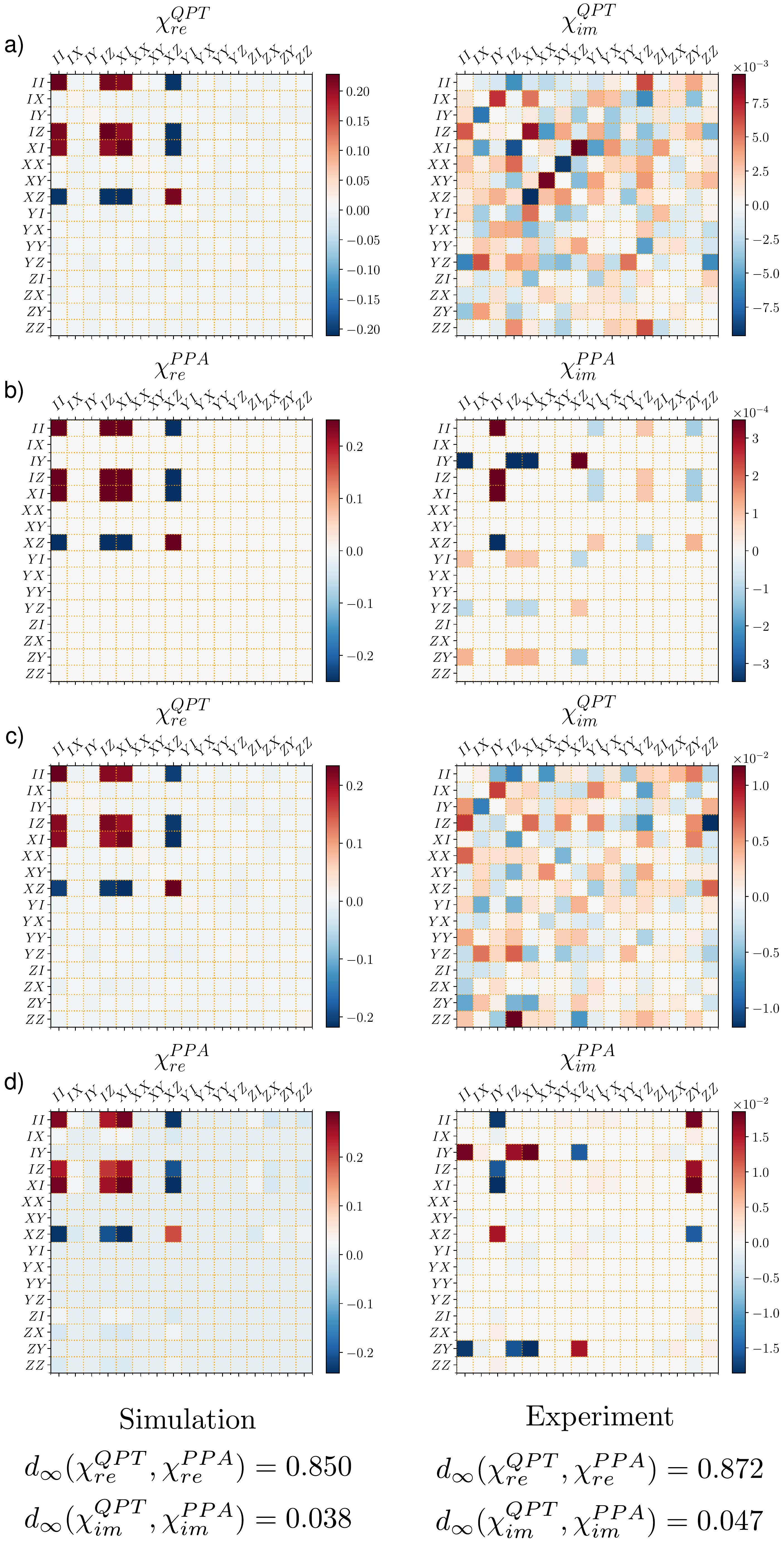} 
    \caption{\emph{$\chi$-matrix for the CX-gate:} Here we depict the $\chi$-matrix representation in Pauli matrices of the CNOT-gate, obtained by the QPT and PPC methods. In a) and b) we used a noiseless simulator to run QPT and PPC, respectively. In c) and d), we used the qubits 0 and 1 on IBMQ-Bogota backend to run QPT and PPC respectively. In the experiments we used $N = 5000$ shots and least squared estimation for the QPT configuration, and $N = 5000$ shots and $N_\theta = 51$ for the PPC configuration. We obtained similar results to the one-qubit case, there is good agreement with the real part, but a slight difference in the imaginary ones. Again, part of the imaginary values stem from an inherent error introduced by the estimation procedure in the QPT method. {In this case, the imaginary part produced by PPC and QPT in the numerical experiments are still statistical no relevant since we can consider the same level of statistical error used in the one-qubit case \cite{Knips2015}, i.e. $\sim 7\times 10^{-3}$. For the physical experiments, the imaginary part overpass the standard deviations ($\sigma_{PPA} \sim 2\times10^{-3}$ and $\sigma_{QPT} \sim 6\times10^{-3}$) reveling imperfections in the quantum gate. In this case, the distance $d_{\infty}(\cdot,\cdot)$ in the simulation is similar with the experimental result. }} 
    \label{fig:cx_char}
\end{figure}

When comparing QPT and PPC-generated process matrices, we observe qualitatively similar results for the real part, $\chi_{re}$, while observing differences in the imaginary part, $\chi_{im}$. In Figures \ref{fig:h_char} and \ref{fig:cx_char}, we present the heat plots of the process matrix from the numerical and \emph{on-hardware} experiments comparing both methods using the gates H and CX as targets, respectively. We further observe that the $\chi^{PPC}_{im}$ differs from $\chi^{QPT}_{im}$, even in the noiseless numerical experiments. The imaginary part, which gives information about the quantum error \cite{Korotkov2013}, in both procedures is slightly different, owing to the assignment error mitigation implemented in PPC. This isolation of SPAM errors is impossible to do under standard QPT. A common way to compare the empirical and the expected operation is through the process fidelity $F_{\chi } = {\rm tr}[\chi \chi_0] / 4^n$, with $\chi_0$ as the noiseless process matrix. In table \ref{tab:Fidelity}, we show the process fidelity values for every experiment. The PPC results show closer values to $1$ than the QPT results in the numerical experiments, where we used a noiseless simulator as the initial test of the method. In the physical experiments, we still observe that $\chi_{PPC}$ has a more minor contribution from the imaginary part than the result from $\chi_{QPT}$. { The numerical results for $\chi^{PPC}_{im}$ and $\chi^{QPT}_{im}$ are irrelevant since the statistical error $N^{-1/2}/2 \sim 7 \times 10^{-3}$\cite{Knips2015} in the projective measurement. For the physical experiment, we observed similar values for one- and two-qubit in the standard deviation ($\sigma_{PPA} \sim 2\times10^{-3}$ and $\sigma_{QPT} \sim 6\times10^{-3}$) after bootstrapping results from 21 experiments using each method in $10^{4}$ resamples. These deviations determine what values are statistically relevant in the experiment. Therefore, the results for one-qubit experiments are closer to an ideal behavior than the results from two-qubit gates. Additionally, we determine the difference between the process matrices $\chi^{PPC}$ and $\chi^{QPT}$ applying the distance $d_{\infty}(\cdot, \cdot)$}  

\renewcommand{\arraystretch}{1.1}
\begin{table}[h!]
    \centering
    \caption{Process fidelity $F_\chi$ of local and non-local gates.  }
    \begin{ruledtabular}
    \begin{tabular}{lccc}
         configuration & gate & $F_{\chi}$ PPC& $F_{\chi}$ QPT \\ 
         \hline
         Numerical &  & 1.0 & 0.99 \\
         IBMQ-Bogota, qubit 0 &  & 0.99 & 0.92  \\ 
         IBMQ-Bogota, qubit 2 &  & 0.99 & 0.95  \\ 
         IBMQ-Santiago, qubit 1 & X & 0.99  & 0.97   \\  
         IBMQ-Santiago, qubit 3 &  & 0.99 & 0.99  \\ 
         IBMQ-Quito, qubit 0 &  & 0.92  & 0.93  \\ 
         IBMQ-Quito, qubit 1 &  & 0.99 & 0.99  \\ 
         IBMQ-Boeblingen, qubit 0 &  & 0.99  & 0.96  \\ 
         IBMQ-Boeblingen, qubit 4 &  & 0.99 & 0.92  \\

         \hline
         Numerical &  & 1.0 & 0.99\\
         IBMQ-Bogota, qubit 0 &  & 0.96 & 0.90  \\ 
         IBMQ-Bogota, qubit 2 &  & 0.99 & 0.95  \\ 
         IBMQ-Santiago, qubit 1 & H & 0.99 & 0.97   \\ 
         IBMQ-Santiago, qubit 3 &  & 0.99 &  0.99 \\ 
         IBMQ-Quito, qubit 0 &  & 0.93  & 0.94  \\ 
         IBMQ-Quito, qubit 1 &  & 0.99 & 0.99  \\ 
         IBMQ-Boeblingen, qubit 0 &  & 0.99  & 0.95  \\ 
         IBMQ-Boeblingen, qubit 4 &  & 0.96 & 0.90  \\
         \hline
         Numerical &  & 1.0 & 0.98 \\
         IBMQ-Bogota, qubits [1,2] &  & 0.97 & 0.75  \\
         IBMQ-Bogota, qubits [0,1] & CX &  0.96 & 0.73   \\
         IBMQ-manhattan, qubits [0,1] &  & 0.99 & 0.88   \\
         IBMQ-manhattan, qubits [11,17] &  & 0.99  & 0.98 \\
         IBMQ-Boeblingen, qubit 0 &  & 0.99  & 0.86  \\ 
         IBMQ-Boeblingen, qubit 4 &  & 0.99 & 0.81  
    \end{tabular}
    \end{ruledtabular}
    \label{tab:Fidelity}
\end{table}

\section{multi-qubit quantum process characterization
\label{sec:characterization}}

Consider the action of a $n$-qubit quantum operator ${\cal U}$ on the state 
\begin{eqnarray}
	\ket{\psi_{k,s}(\theta)} &=& Y_\theta^s \ket{k_0, \cdots ,k_{s-1}, 0,k_{s+1}, \cdots ,k_{n}} \nonumber \\
	&=& Y_\theta^s \ket{k,0_s} .
\end{eqnarray}
Above, we used a short notation to identify where the rotation is being applied, in this case on the ground state of the $s$-th qubit while the rest are at $\ket{k_0, \cdots ,k_{s-1},k_{s+1}, \cdots ,k_{n}}$. The probability distribution for ${\cal U} \ket{\psi_{k,s}(\theta)}$ reads as 
\begin{equation}\label{eq:Umodel}
	P^{\, \cal U}_{k,s} (\theta) = \frac{1}{2} \left( A_{k,s} + B_{k,s} \, c_{\theta} + C_{k,s} \, s_\theta \right) , 
\end{equation}
where $A_{k,s}$, $B_{k,s}$, and $C_{k,s}$ are vectors in $\mathbb{R}^{2^n}$, with components 
\begin{eqnarray}
	[A_{k,s}]_j &=& |\bra{j}{\cal U} \ket{k,0_s}|^2 + |\bra{j}{\cal U} \ket{k,1_s}|^2 \ , \\   						 
	\, [B_{k,s}]_j &=& |\bra{j}{\cal U} \ket{k,0_s}|^2 - |\bra{j}{\cal U} \ket{k,1_s}|^2 \ , \\
	\, [C_{k,s}]_j &=& \bra{j}{\cal U} \ket{k,0_s} \bra{j}{\cal U} \ket{k,1_s}^* + \nonumber \\
	                             && \quad \quad \bra{j}{\cal U} \ket{k,0_s}^* \bra{j}{\cal U} \ket{k,1_s} , \  j \in \{0,1\}^n \ . \nonumber \\
\end{eqnarray}
By measuring $N$ systems identically prepared in the state ${\cal U} \ket{\psi_{k,s}(\theta)}$ for every $\theta$ in ${\cal S}_{N_\theta}= \{(j/N_\theta - 1)\pi : j\in \mathbb{N}, 0 \leq j \leq N_\theta \}$; we estimate the probability distributions $Q_{k,s}^j = ({\rm C}^{k,j}_0/N, \cdots, {\rm C}^{k,j}_{2^n}/N)$ for $j = 0, \cdots, N_\theta$, with ${\rm C}^{k,j}_m$ as the number of outcomes `$m$' $\in \{0, 1 \}^n$ for the $j$-th angle. The original distributions $P^j_{k,s}$, after mitigating the assignment error (see next section for details), are obtained by minimizing the functions $f_j(P) = ||Q^j_{k,s} - T\cdot P||_F$, where $||\cdot ||_F$ is the Frobenius norm. With that information, i.e.~$\{ P^0_{k,s}, \cdots, P^{N_\theta}_{k,s} \}$ and the model \eqref{eq:Umodel}, we can compute the matrix elements for $\cal U$. Therefore, we can determine the process matrix $\chi$.

\section{calibration
\label{sec:calibration}}
We assume the assignment error effects are represented by a transition matrix $T_{kl} = P(l|k)$, where $P(l|k)$ is the conditional probability of observing $\ket{j}$ when the system has been prepared in the state $\ket{i}$. This is typically determined by measuring the number of outcomes `$j$', $c_j$, of $N$ identically-prepared states of $\ket{i}$; $T_{ij} = c_j/N$. However, this procedure is limited in its explanatory power, as we cannot determine which part of the readout error comes from the state preparation. Additionally, this method scales exponentially with the number of qubits considered since we must evaluate the conditional probabilities for all $2^n$ states of the system. We follow a modified calibration routine that tracks assignment error as a function of the rotation about the $y$-axis to discriminate the assignment error from every possible state preparation error in each experiment.

We rotate the states $\ket{0,k_0, \cdots, k_{n-1}}$ around the $y$-axis of the first qubit, for $k \in \lbrace 0,1 \rbrace^{n-1}$, yielding the output $c_{\theta/2} \ket{0, k_0, \cdots, k_{n-1}} + s_{\theta/2} \ket{1, k_0, \cdots, k_{n-1}}$. Thus, the expected probability distribution will have two components, $c_{\theta/2}^2$ and $s_{\theta/2}^2$, for instance 
\begin{eqnarray}
	P_{k = 0\cdots 0}(\theta) &=& (c_{\theta/2}^2,\underbrace{0,\cdots,0}_{2^{n-1}-1},s_{\theta/2}^2, 0,\cdots,0)^T , \nonumber \\
    P_{k = 10\cdots 0}(\theta) &=& (0, c_{\theta/2}^2,\underbrace{0,\cdots,0}_{2^{n-1}-2},0,s_{\theta/2}^2,0,\cdots,0)^T , \nonumber \\
    & \vdots & \nonumber \\
    P_{k = 0\cdots 01}(\theta) &=& (\underbrace{0,\cdots,0}_{2^{n-1}-1}, c_{\theta/2}^2,0,\cdots,0,s_{\theta/2}^2)^T . \nonumber \\
\end{eqnarray} 
Under the bit-flip noise model, an empirical probability distribution $Q_{k}(\theta)$ is related to $P_{k}(\theta)$ by $Q_{k}(\theta) = T \cdot P_{k}(\theta)$, with 
\begin{equation}\label{eq:Tmodel}
	T = \left( \begin{array}{ccc}
		t_{00} & \cdots & t_{0,2^n} \\
		\vdots & \ddots & \vdots \\
		t_{2^n,0} & \cdots & t_{2^n,2^n}
	\end{array}
	\right) \ .
\end{equation}
Taking into account the structure of $P_{k}(\theta)$, we can determine two columns of $T$ per state $Y_\theta \ket{0,k_0, \cdots, k_{n-1}}$. We measure the states $Y_{\theta_j} |0, k_0, \cdots, k_{n-1} \rangle $, $\theta_j \in {\cal S}_\theta$, and estimate the probability distribution by the sampling distribution of the outcomes, i.e.~we obtain the quantities $Q_k^j = ({\rm C}^{k,j}_0/N, \cdots, {\rm C}^{k,j}_{2^n}/N)^T$. We fit the acquired data with the model $T \cdot P_k (\theta - \theta_0)$. In the model, we have considered an initial phase $\theta_0$ to describe preparation error.

\subsection{Two-qubit transition matrix and initial preparation}
For the two-qubit system, we can consider the following model for the expected distributions of the states $Y^0_\theta \ket{00}$ and $Y^0_\theta \ket{01}$,  
\begin{eqnarray}
    P_{0}(\theta) &=& (c_{\theta/2}^2,0,s_{\theta/2}^2, 0)^T , \nonumber \\
    P_{1}(\theta) &=& (0, c_{\theta/2}^2,0,s_{\theta/2}^2)^T , 
\end{eqnarray}
and the emperical distributions $Q_k(\theta) = T \cdot P_k(\theta)$, 
\begin{eqnarray}\label{eq:twoqubitmodel}
    Q_{0}(\theta) &=& \left( \begin{array}{c}
        t_{00} \\
        t_{10} \\
        t_{20} \\
        t_{30} 
    \end{array} \right) c_{\theta/2}^2 +  \left( \begin{array}{c}
        t_{02} \\
        t_{12} \\
        t_{22} \\
        t_{32} 
    \end{array} \right) s_{\theta/2}^2  , \nonumber \\
    Q_{1}(\theta) &=& \left( \begin{array}{c}
        t_{01} \\
        t_{11} \\
        t_{21} \\
        t_{31} 
    \end{array} \right) c_{\theta/2}^2 +  \left( \begin{array}{c}
        t_{03} \\
        t_{13} \\
        t_{23} \\
        t_{33} 
    \end{array} \right) s_{\theta/2}^2  . \nonumber \\ 
\end{eqnarray}
Now, we execute the quantum circuits
\begin{eqnarray}
    \Qcircuit @C=1.4em @R=1.8em {
   \lstick{\ket{0}} & \gate{Y_{\theta_j}} & \meter \\
   \lstick{\ket{0}} & \gate{\mathbb{I}} & \meter
  } \quad {\rm for}\ k=0 , \\[4mm]
  \Qcircuit @C=1.4em @R=1.8em {
   \lstick{\ket{0}} & \gate{Y_{\theta_j}} & \meter \\
   \lstick{\ket{0}} & \gate{X} & \meter
  } \quad {\rm for}\ k=1 ,
\end{eqnarray}
and for $\theta_j \in {\cal S}_\theta$. We determine the transition matrix elements $t_{ij}$ by fitting the model \eqref{eq:twoqubitmodel} to the acquired data $Q_k^j = ({\rm C}_{00}^{k,j}/N, {\rm C}_{01}^{k,j}/N,{\rm C}_{10}^{k,j}/N,{\rm C}_{11}^{k,j}/N)^T$ from the above quantum circuit executions. 
 
\section{Discussion} 
To summarize, we introduced an alternative method to characterize a {unitary} quantum process based on a rotational sweeping procedure {without prior information about the process}. Instead of using a complete set of measurement operators, we measure the unknown quantum state rotated around different angles. Additionally, we proposed a pre-characterization process to define the assignment errors enclosed in the transition matrix and then mitigated those effects in the quantum process characterization. In the scheme presented here, we perform $2^{n-1}(n+2) N_\theta$ experiments to characterize an  $n$-qubit process, in which the number of angles $N_\theta$ does not scale with the number of qubits. Therefore, we can fix this parameter to get enough data to fit the model with minor deviations.  We considered $N_\theta = 51$ for a suitable fitting with relatively slight deviations ($\sim 10^{-4}$) from ideal behavior; this value can be used for any $n$-qubit system. Additionally, QPT is affected by errors in preparing the initial states and tomography measurements. By contrast, in our method, we control the initial preparation error by considering an initial phase in the model and the readout error by mitigating the assignment error.     

The PPC algorithm is affected by the fitting error that depends on the number of rotational divisions, $N_\theta$. This procedure assumes a high-fidelity rotation and trusted quantum states $\ket{0}$ and $\ket{1}$ in the same way that QPT relies on high fidelity measurement. One way to mitigate a possible imperfection in the rotation process is by using a different compilation for the rotations $Y_\theta$ and $X_\theta$. Instead of decomposing any rotation in terms of $X_{\pm \pi /2}$ and $Z_\theta$, we can use the compilation procedure introduced in \cite{Gokhale2020}; a decomposition in terms of $X_\theta$. The alternative decomposition brings a shorter pulse structure and commensurately higher fidelity. 

To validate the performance of the PPC procedure, we compared the process matrix obtained by our approach with the outcome from the QPT. We compared the process fidelity $F_\chi$ of the results, from PPC and QPT, of different one- and two-qubits gates {\it on-hardware} and {\it in-silico}. As expected, we found minor differences between ${\rm Re}\lbrace \chi_{PPC} \rbrace$ and ${\rm Re}\lbrace \chi_{QPT} \rbrace$; there was, however, a remarkable difference in their imaginary parts, even in the numerical experiments. {The matrix elements $(\chi^{QPT/PPA}_{im})_{ij}$ can be attributed to numerical errors. They are in the range of the statistical error present in the simulation and therefore do not contribute to the characterization analysis. We observed statistically relevant values for $\chi^{QPT}_{im}$ in the physical experiments and neglishible ones in $\chi^{PPA}_{im}$, other consequence of the SPAM mitigation in the PPA process in one-qubit quantum processes. The PPA implements the mitigation of preparation errors in local rotations, therefore the imaginary part $\chi^{QPT}_{im}$ contents relevant data.  We establish the difference of the process matrices via the distance $d_{\infty}(\cdot, \cdot)$, and the process fidelity as performance metric that gives a perfect score in the noiseless simulation for the PPA (see Table \ref{tab:Fidelity}). }

Additionally, the PPC protocol allows the calculation of the error process matrix directly from the quantum gate's characterization without appealing to the QPT process introduced by Korotkov \cite{Korotkov2013}. The imaginary part of the error process matrix provides information about the process fidelity and imperfections. A natural extension of this work is studying the error process matrix for low-depth quantum circuits. The method may also prove useful in providing tighter measures of crosstalk effects in quantum processes. For example, tomography on one qubit, while its neighbors undergo local unitaries, can reveal correlated noise~\cite{idle1,idle2}. The improved scalability of PPC over QPT allows us to extend this tomographic method to more significant numbers of qubits, a valuable feature for crosstalk identification and characterization.

\appendix

\section{Quantum process tomography implementation}\label{app:qpt}
The QPT algorithm finds the process matrix $\chi_\mathcal{E}$ of a quantum map $\mathcal{E}: \kket{\rho} \rightarrow \kket{E \rho E^\dagger}$, by measuring the resulting state $\mathcal{E}\kket{\rho_k}$, from a basis of initial states $\mathcal{P}: \lbrace \kket{\rho_1}, \cdots, \kket{\rho_{4^n}}\rbrace$, onto different directions $\mathcal{M}:\lbrace \kket{E_1}, \cdots , \kket{E_M} \rbrace$ in the Hilbert space, with $n$ and $M$ as the number of qubits and number of measurement operators respectively. Here, we have introduced the superoperator notation for the statistical operator, where operators become superkets and quantum maps becomes superoperators, i.e, $\mathcal{E}(\rho) \rightarrow \mathcal{E}\kket{\rho}$ (more details in \cite{Greenbaum2015}). In the superoperator notation, the goal is to determine the matrix representation $[ \chi_\mathcal{E} ]_{ij} = \bbra{j}| \mathcal{E} | \kket{i}$ in the Pauli basis $\lbrace \kket{i} \rbrace$, by the measurements 
\begin{equation}\label{eq:lambda}
    \lambda_{ij} = \bbra{E_j} \mathcal{E} \kket{\rho_i}. 
\end{equation}  
We can establish the relation between $\lambda$ and $\chi_\mathcal{E}$ by inserting the completeness identity $\sum_i \kket{i}\bbra{i} = \mathbb{I}$ in Eq.~\ref{eq:lambda},

\begin{equation}\label{eq:chi1}
    \lambda_{ij} = \sum_{k,l = 1}^{4^n}  [\chi_\mathcal{E}]_{lk} \ \bbra{E_j} k \rangle \rangle \langle \langle l \kket{\rho_i} .
\end{equation}
We can arrange the terms as 
\begin{eqnarray}
    y_{j + (i-1)\times M} &=& \lambda_{ij}, \\
    x_{k + (l-1)\times 4^n} &=& [\chi_\mathcal{E}]_{ij} \\
    B_{j + (i-1)\times M, \, k + (l-1)\times 4^n} &=&  \bbra{E_j} k \rangle \rangle \langle \langle l \kket{\rho_i}, 
\end{eqnarray}
and transform \ref{eq:chi1} into a more convenient expression, $B \, \vec{x} - \vec{y} = 0 $, for a numerical solution.  

\begin{figure}
    \centering
    \includegraphics[width=\linewidth]{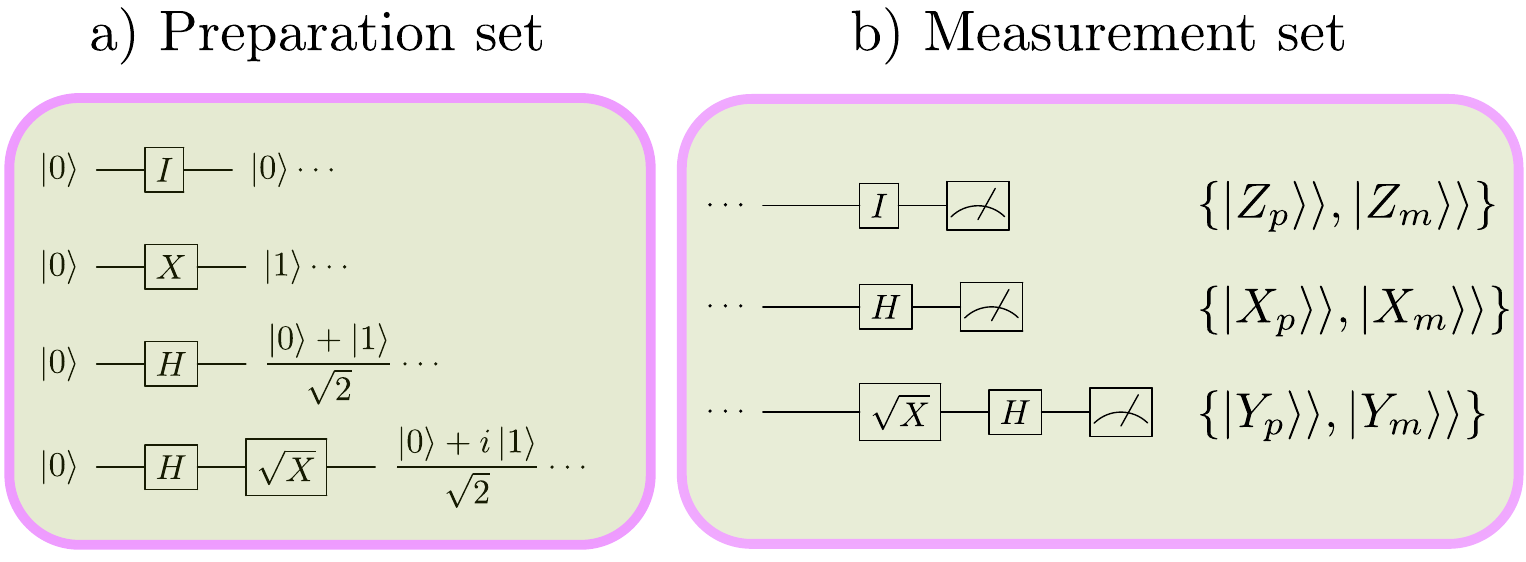}
    \caption{Preparation and Measurement set of quantum circuits for one-qubit quantum process tomography.} 
    \label{fig:onequbitqpt}
\end{figure}

{\it B.1 One-qubit process matrix:} Without loss of generality, consider the characterization of a one-qubit quantum map $\mathcal{E}$ using the following intitial states and measurement operators
\begin{equation}\label{eq:prepmeas}
    \begin{aligned}[c]
        \mathcal{P} :  \left\lbrace \right. \, &  \kket{Z_p},  \kket{Z_m} , \\
                         & \kket{X_p}, \kket{Y_p} \, \left. \right\rbrace \\ &
    \end{aligned} 
    \quad 
    ,
    \quad
    \begin{aligned}[c]
        \mathcal{M} :  \left\lbrace \right. \, &  \kket{Z_p},  \kket{Z_m} , \\  
                         & \kket{X_p},  \kket{X_m}, \\
                         &  \kket{Y_p},  \kket{Y_m} \, \left. \right\rbrace  \ \ ,    
    \end{aligned} 
\end{equation}
where we have introduced the projectors $Z_p = (\mathbb{I} + \sigma_z)/2$, $Z_m =(\mathbb{I} - \sigma_z)/2$, $X_{p/m} = Y_{-\pi/2} Z_{p/m} Y_{\pi/2}$, and $Y_{p/m} = X_{-\pi/2} Z_{p/m} X_{\pi/2}$.  Since $Z_p = \ket{0}\bra{0}$, $Z_m = \ket{1}\bra{1}$, $X_p = \ket{+}\bra{+}$, and $Y_p = \ket{i}\bra{i}$, the set $\mathcal{P}$ is generated by the preparation of the states $\lbrace \ket{0}, \ket{1}, \ket{+} , \ket{i} \rbrace$. The quantum chip detector measures $\sigma_z$ by default. The possible outcomes of a measurement on an arbitrary state $\rho$ are $i = 0, 1$, where $i=0$ corresponds to $\langle \sigma_z \rangle  = +1$ and $i = 1$ to $\langle \sigma_z \rangle = -1$, with probabilities
\begin{equation}
    p_0 = {\rm tr} \lbrace \rho Z_p   \rbrace, \quad p_1 = {\rm tr} \lbrace \rho Z_m   \rbrace.  
\end{equation}     
Now, to measure $\sigma_x$ and $\sigma_y$ we need to consider the projectors $X_p$, $X_m$, and $Y_p$, $Y_m$, respectively, by transforming the $x$- and $y$-axis into the $z$-axis in the Bloch sphere and measuring $\sigma_z$ (see the definitions below Eq.~\ref{eq:prepmeas}). Therefore, the set $\mathcal{M}$ is generated by the gates $\lbrace \mathbb{I}, Y_{-\pi/2}, X_{-\pi/2} \rbrace$. In Figure \ref{fig:onequbitqpt} there is a sketch of the required circuits to gate characterization. Thus, the number of quantum circuits for a complete characterization is less than the number of independent terms in the matrix process $\chi_{\mathcal{E}}$. On the other hand, one can observe that the number of measurement operators, size of $\mathcal{M}$, is enough for the solution of Eq.~\ref{eq:lambda}, since ${\rm dim} \lbrace \mathcal{M}\rbrace\times {\rm dim} \lbrace \mathcal{P}\rbrace > {\rm dim} \lbrace \vec{y} \rbrace$.

{
{\it B.2 $n$-qubit process matrix:} The natural extension of the initial states and the measurement operators in the $n$-qubit quantum process characterization follows:
\begin{equation}\label{eq:prepmeas}
    \begin{aligned}[c]
        \mathcal{P}_{n} :  \left\lbrace \right. \, &  \kket{Z_p},  \kket{Z_m} , \\
                         & \kket{X_p}, \kket{Y_p} \, \left. \right\rbrace^{\otimes n} \\ &
    \end{aligned} 
    \quad 
    ,
    \quad
    \begin{aligned}[c]
        \mathcal{M}_{n} :  \left\lbrace \right. \, &  \kket{Z_p},  \kket{Z_m} , \\  
                         & \kket{X_p},  \kket{X_m}, \\
                         &  \kket{Y_p},  \kket{Y_m} \, \left. \right\rbrace^{\otimes n}  \ \ ,    
    \end{aligned} 
\end{equation}
where $\mathcal{P}_{n}$ is generated by the preparation preparation of the states $\lbrace \ket{0}, \ket{1}, \ket{+} , \ket{i} \rbrace^{\otimes n}$. Now, the possible outcomes of the $\sigma_{z}^{\otimes n}$ measurement on an arbitrary state are $s = \lbrace 0,1 \rbrace^{n}$, with probabilities 
\begin{eqnarray}
	p_{0\cdots0} &=& {\rm tr} \lbrace Z_{p}\otimes \cdots \otimes Z_{p} \rbrace \nonumber \\
	p_{010\cdots0} &=& {\rm tr} \lbrace Z_{p}\otimes Z_{m} \otimes  Z_{p} \otimes \cdots \otimes Z_{p} \rbrace \nonumber \\
	                       &\vdots & \nonumber \\
	p_{1\cdots1} &=& {\rm tr} \lbrace Z_{m}\otimes \cdots \otimes Z_{m} \rbrace  .                 
\end{eqnarray}
For the $\lbrace \sigma_{z}, \sigma_{y}, \sigma_{z} \rbrace^{\otimes n}$ measurements we need to apply the projectors $\lbrace X_{p} , X_{n}\rbrace^{\otimes n}$ and $\lbrace Y_{p} , Y_{n}\rbrace^{\otimes n}$, respectively. Thus, the set $\mathcal{M}_{n}$ is generated by $\lbrace \mathbb{I}, Y_{-\pi/2}, X_{-\pi/2} \rbrace^{\otimes n}$. Again, as in the one-qubit case, we get a redundant amount of measurements, ${\rm dim} \lbrace {\cal M}_{n}\rbrace \times {\rm dim} \lbrace {\cal P}_{n}\rbrace = 4^{n} \times 6^{n}$ for the number $16^{n}$ of independent elements in $\chi$. 
}
\section{Fitting parameters} \label{app:exp}
In this section we shall discuss the fitting details used in the post-processing step in the PPC protocol. 

\begin{widetext}

\begin{figure}[h]
    \centering
    \includegraphics[width=0.6\linewidth]{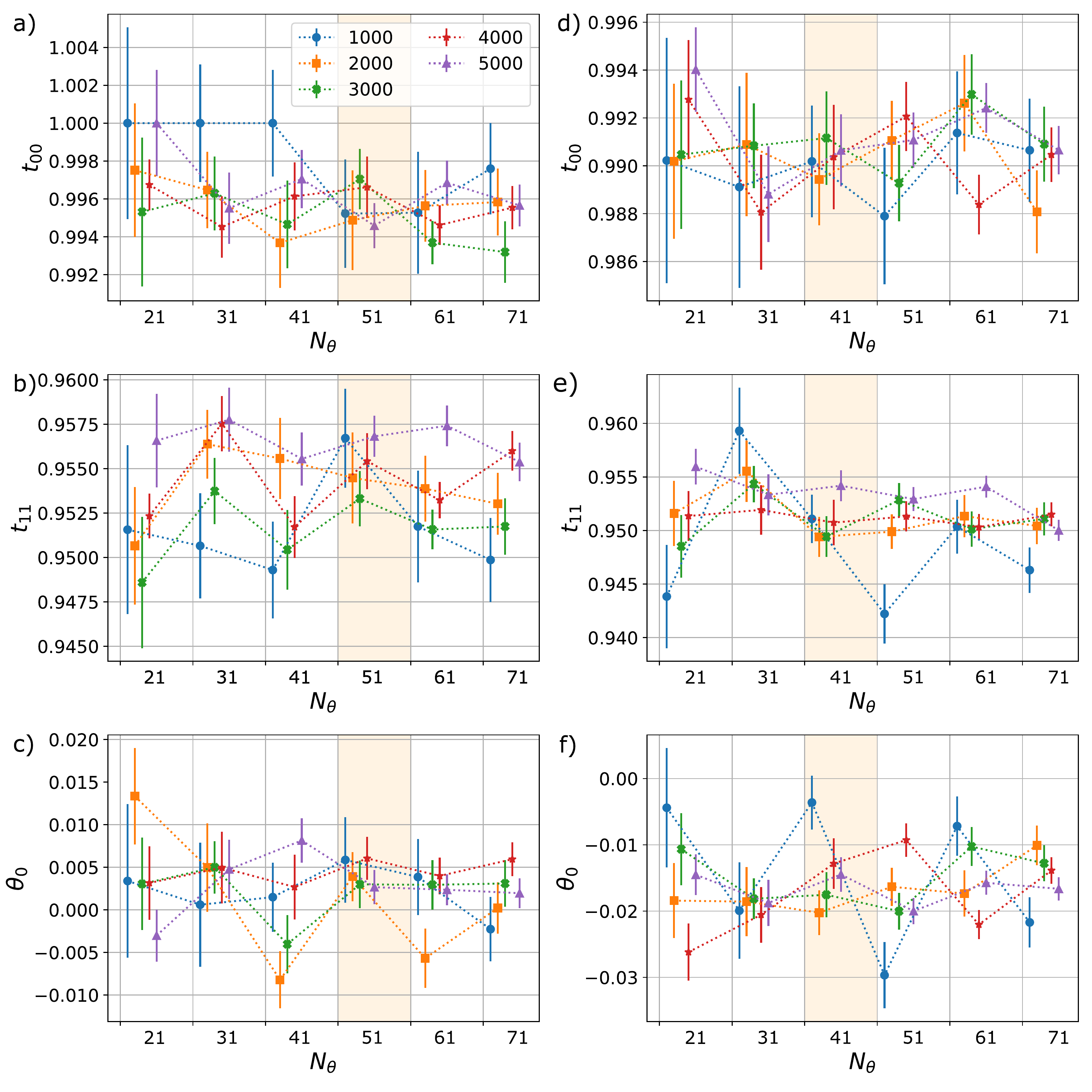}
    \caption{ \emph{Fitting parameters:} Conditional probabilities and initial phase using $Y_\theta$ rotation in a), b), and c), and using the $X_\theta$ rotation in d), e), and f), respectively. The error bars represent the standard deviation of each parameter in the fitting process. The shadow regions indicate the optimal values for $N_\theta$ and $N$.}
    \label{fig:cal}
\end{figure}

\end{widetext}

The number of rotations and shots plays an essential role in the quantum process characterization. We consider a one-qubit experiment to benchmark the calibration process, which follows the same principle as the characterization. In this experiment, we consider the qubit 0 in the IBMQ-Bogota quantum chip. For the calibration we use rotations about the $x$- and $y$-axis, with a different number of rotations and shots. Figure \ref{fig:cal} shows the conditional probabilities and the initial phase for each experimental setup as a function of the number of rotations and shots, with error bars indicating the standard deviation in each measurement. We choose a suitable setup where the conditional probabilities do not vary significantly concerning the result using the highest values $N_\theta = 71$ and $N = 5000$. For $Y_\theta$ and $X_\theta$ we found the optimal points $N_\theta = 51$ and $N_\theta = 41$, respectively (see shadow regions in Figure \ref{fig:cal}). One important feature is the dependence of the parameters' standard deviation on $N_\theta$ and $N$, which slightly improves the experimental setup's refinement. 

\begin{acknowledgments}
    We thank Ryan Bennink for valuable discussions.  This work was supported as part of the ASCR Quantum Testbed Pathfinder Program at Oak Ridge National Laboratory under FWP \# ERKJ332. This research used resources of the Oak Ridge Leadership Computing Facility, which is a DOE Office of Science User Facility supported under Contract DE-AC05-00OR22725.
    \end{acknowledgments}
    
    \section*{Author contributions}
    V.L.O. introduced the PPC method. V.L.O. and T.K. developed the theoretical formalism, performed the analytic calculations and performed the experiments. R.C.P.~analyzed and interpreted experimental data and suggested experiments to run. V.L.O, T.K., and R.C.P. cowrote the manuscript.
    \section*{Competing interests}
    The authors declare that there are no competing interests.

\end{document}